\documentclass[
  reprint,            
  amsmath,amssymb,    
  aps,                
  prx,                
  floatfix,            
  superscriptaddress
]{revtex4-2}

\usepackage{physics}
\usepackage[T1]{fontenc}

\usepackage{amsmath}
\usepackage{amsfonts}
\usepackage{amssymb}
\usepackage{bm}
\usepackage{mathrsfs}

\usepackage{graphicx}
\usepackage{subcaption}
\usepackage{float}
\usepackage{indentfirst}
\usepackage{booktabs}
\usepackage{xcolor}
\usepackage[colorlinks=true,
            citecolor=blue,
            linkcolor=blue,
            urlcolor=blue,
            filecolor=blue]{hyperref}

\begin{document}

\title{\bf {Physics-Informed Neural Quantum Control for Rovibrational Photoassociation in a Morse Molecular System}}

\author{Murilo D. Forlevesi}
\email{murilo.deliberali@unesp.br }
\author{Edson Denis Leonel}
\email{edson-denis.leonel@unesp.br }
\affiliation{Departamento de Física, UNESP - Universidade Estadual Paulista, Rio Claro SP, 13506-900, Brazil}
\author{Emanuel Fernandes de Lima}
\email{eflima@ufscar.br}
\affiliation{Departamento de Física, Universidade Federal de São Carlos (UFSCar)\\ São Carlos, SP 13565-905, Brazil}

\date{\today}

\begin{abstract}

We present a Physics-Informed Neural Quantum Control (PINQC) framework for rovibrational photoassociation in a Morse molecular system. The proposed method combines neural-network-based laser-field generation with differentiable quantum propagation, allowing optimized laser pulses to be obtained directly from the underlying quantum dynamics. The optimized control fields efficiently transfer an initially Gaussian wave packet into the vibrational ground-state level, promoting continuum-to-bound population transfer through coherent rovibrational dynamics. The resulting quantum dynamics involve multiple physical processes, including continuum-to-bound photoassociation, vibrational stabilization, and rotational redistribution arising from dipole-induced couplings between neighboring rotational levels. A central result of the present work is the successful application of the PINQC framework to rovibrational models containing larger rotational levels than those previously accessible in our conventional optimization calculations. The optimization remains numerically stable despite the increased complexity of the molecular system, demonstrating that differentiable optimization provides an effective strategy for treating rovibrational models of increased dimensionality. These results establish the PINQC framework as a promising computational tool for molecular photoassociation and motivate future investigations of increasingly complex rovibrational quantum-control problems.

\end{abstract}
\maketitle
\newpage

\section{Introduction}

The coherent manipulation of quantum systems by externally applied electromagnetic fields has become one of the central topics in modern atomic, molecular, and optical physics, with broad applications ranging from molecular dynamics and ultracold chemistry to quantum information processing and emerging quantum technologies \cite{Dong2010,dAlessandro2021,Tannor1985control,Shapiro1986,Tannor1986coherent}. Since the pioneering ideas of coherent control were introduced, Quantum Optimal Control (QOC) has evolved into a mature interdisciplinary field connecting quantum mechanics, optimal control theory, numerical optimization, and computational physics \cite{Rabitz2000,Glaser2015,Ansel2024,Botina1996,Brif2010,Rabitz1990,Londono2023}.

Among the many applications of quantum control, molecular photoassociation represents one of the most demanding optimization problems. In this process, two initially unbound atoms must be coherently transferred into deeply bound molecular states through appropriately tailored laser fields \cite{Jones2006,Koch2019,Gacesa2013,Lyu2019,Devolder2018proposal,deLima2024formation,DeLima2017,Lyu2019,ndong2010}. Achieving this objective generally requires simultaneously controlling several physical mechanisms, including long-range capture, vibrational stabilization, rotational redistribution, and interference between multiple quantum pathways. Consequently, photoassociation naturally gives rise to high-dimensional quantum-control problems involving strongly coupled rovibrational dynamics \cite{Shapiro2012,Chakrabarti2007,Horiba2022}.

Over the past three decades, a variety of quantum optimal-control algorithms have been developed to address problems of this nature. Variational methods based on Krotov's formulation, gradient-based optimization algorithms such as GRAPE, spectral approaches including CRAB, and iterative procedures such as the Two-Point Boundary-Value Quantum Control Paradigm (TBQCP) have demonstrated remarkable success in molecular dynamics, coherent spectroscopy, quantum gates, and ultracold molecular formation \cite{Krotov1996,Zhu1998,Khaneja2005,Caneva2011}. Nevertheless, as molecular models evolve toward increasingly realistic rovibrational descriptions, extending these optimization procedures to larger Hilbert spaces becomes progressively more demanding from a computational standpoint.

Despite their remarkable success, extending conventional quantum optimal-control calculations to increasingly realistic rovibrational models requires substantially higher computational resources because each optimization iteration involves repeated forward and backward propagations of the time-dependent Schrödinger equation \cite{Koch2022}. As additional rotational levels are incorporated into the molecular model, both the memory requirements and execution times increase significantly, often limiting the practical size of the systems that can be investigated \cite{Shimshon2022}.

This limitation became particularly evident in our previous study on rovibrational photoassociation using the TBQCP framework \cite{ForlevesiTBQCP}. Although efficient continuum-to-bound population transfer was achieved for systems containing a moderate number of rotational levels, extending the calculations to larger rotational levels rapidly became computationally prohibitive due to the increasing memory requirements and optimization cost. These computational limitations motivated the search for alternative optimization strategies capable of treating larger rovibrational models while preserving the underlying quantum dynamics.

Recent advances in Scientific Machine Learning (SciML) have established differentiable programming as a powerful computational paradigm by integrating physical models, automatic differentiation, and machine learning into a single differentiable framework \cite{rackauckas2020universal,Karniadakis2021,hang2024unisolver,hao2024dpot}. In this context, physics-informed methods constitute one important class of SciML approaches, in which the governing physical equations are incorporated directly into the optimization process rather than learned from data \cite{Raissi2019,Cuomo2022,Liu2026,liu2026data}. These developments have created new opportunities for quantum-control optimization by enabling gradients to propagate through differentiable physical models \cite{Chen2018,Innes2019,nieves2024}, motivating the investigation of differentiable optimization strategies for molecular quantum control.

In this work, we introduce a Physics-Informed Neural Quantum Control (PINQC) framework for rovibrational photoassociation in a molecular Morse potential. The proposed method combines a neural-network parametrization of the control field with differentiable quantum propagation, allowing laser pulses to be optimized directly from the underlying quantum dynamics. The principal motivation of the present work is to investigate whether this differentiable formulation enables the optimization of molecular photoassociation problems involving larger rovibrational models than those that could be treated in our previous TBQCP-based optimization of the photoassociation process, where practical computational limitations restricted the calculations to smaller rotational manifolds. We demonstrate that the proposed PINQC framework successfully extends these previous calculations to larger rotational levels while preserving the underlying quantum dynamics, stable convergence, and efficient continuum-to-bound population transfer. These results establish PINQC as a practical framework for investigating increasingly complex rovibrational quantum-control problems involving larger rotational manifolds.

\section{Rovibrational Model and Physics-Informed Neural Quantum Control}

\subsection{Rovibrational model}

The present work employs the same rovibrational molecular model introduced in our previous study \cite{ForlevesiTBQCP}, where its complete derivation and numerical implementation are described in detail. Here, we briefly summarize the main ingredients required for the development of the proposed differentiable optimization framework.

The photoassociation process is formulated as a rovibrational quantum optimal-control problem in which an external laser pulse drives an initially continuum-like wave packet into the vibrational ground-state manifold of a molecular Morse potential \cite{DeLima2011,DeLima2015,Ulmanis2012,Kerman2004,Zhang2020}. The molecular dynamics is governed by the time-dependent Schrödinger equation

\begin{equation}
i\hbar\frac{\partial}{\partial t}\Psi(r,\theta,\phi,t)
=
\left[
\hat{H}_0
+
\hat{H}_1(t)
\right]
\Psi(r,\theta,\phi,t),
\end{equation}
where the field-free Hamiltonian is
\begin{equation}
\hat{H}_0
=
-
\frac{\hbar^2}{2\mu}\nabla^2
+
V_M(r),
\end{equation}
with $V_M(r)$ described by the Morse potential
\begin{equation}
V_M(r)
=
D_e
\left(
1-e^{-a(r-r_e)}
\right)^2,
\end{equation}
where $D_e$, $a$, and $r_e$ denote the dissociation energy, the potential width parameter, and the equilibrium internuclear distance, respectively. The interaction with the external laser field is described within the electric-dipole approximation\cite{Korolkov1996},

\begin{equation}
\hat{H}_1(t)
=
-\mu(r)E(t)\cos\theta,
\end{equation}
where $E(t)$ is the control field and the molecular transition dipole moment is modeled as
\begin{equation}
\mu(r)
=
q\,r\,e^{-r/r_d}.
\end{equation}

Following the procedure described in Ref.~\cite{ForlevesiTBQCP}, the Schrödinger equation is projected onto a truncated rovibrational basis composed of bound and discretized continuum eigenstates. For each rotational quantum number $l$, all bound vibrational states together with a finite number of box-normalized continuum states are retained, yielding a finite-dimensional Hilbert space suitable for numerical propagation. Throughout this work, only the subspace with magnetic quantum number $m=0$ is considered, consistent with the use of a linearly polarized laser field. The basis dimension is subsequently varied to investigate the capability of the proposed PINQC framework to optimize increasingly complex rovibrational systems. Owing to the angular dependence of the dipole interaction, the rotational selection rule

\begin{equation}
\Delta l=\pm1,
\end{equation}
which couples neighboring rotational levels during the photoassociation dynamics. The initial state is taken as a continuum-like Gaussian wave packet,
\begin{equation}
\Psi(r,\theta,\phi,0)
=
\frac{\xi(r)}{r}
Y_0^0(\theta,\phi),
\end{equation}
where
\begin{equation}
\xi(r)=
\left(
\frac{2}{\pi d^2}
\right)^{1/4}
\exp
\left[
ik_0r
-
\frac{(r-r_0)^2}{d^2}
\right].
\end{equation}

The corresponding expansion coefficients are obtained by projecting the initial wave packet onto the rovibrational basis described above. The optimization objective is to maximize the population transferred to the vibrational ground-state manifold,

\begin{equation}
F
=
\sum_l
\left|
\langle
v=0,l
|
\Psi(t_f)
\rangle
\right|^2,
\end{equation}
which corresponds to the total population accumulated in the vibrational ground state independently of the final rotational quantum number. Consequently, the optimization explicitly targets vibrational stabilization, while the final rotational population distribution emerges naturally from the optimization process.

\subsection{Physics-Informed Neural Quantum Control}

To optimize the laser pulse associated with the photoassociation process, we introduce a Physics-Informed Neural Quantum Control (PINQC) framework inspired by recent advances in physics-informed machine learning and differentiable programming \cite{gao2021,geneva2020,ren2022,zhu2019}. The proposed approach parametrizes the control field directly through a neural network while preserving the exact quantum dynamics through an explicit differentiable Schrödinger propagator. Consequently, gradients are propagated through the complete quantum evolution by automatic differentiation, enabling end-to-end optimization of the control field. The control field is generated according to
\begin{equation}
E(t)=\pi_{\boldsymbol{\theta}}(t),
\end{equation}
where $\pi_{\boldsymbol{\theta}}$ denotes a neural policy parametrized by the trainable parameter vector $\boldsymbol{\theta}$. The neural network therefore acts as a continuous control policy mapping the time coordinate into the instantaneous electric-field amplitude.

Instead of using the raw time coordinate directly, the normalized time variable,
\begin{equation}
\tau=\frac{t}{T},
\qquad
\tau\in[0,1],
\end{equation}
is transformed through a Fourier-feature embedding composed of sinusoidal basis functions \cite{Tancik2020,Sitzmann2020},
\begin{equation}
\begin{aligned}
\mathbf{x}(\tau)
={}&
\Big[
\tau,
\sin(2\pi\omega_1\tau),
\cos(2\pi\omega_1\tau),\\
&\qquad
\dots,
\sin(2\pi\omega_N\tau),
\cos(2\pi\omega_N\tau)
\Big].
\end{aligned}
\end{equation}

This representation mitigates the spectral bias of neural networks and facilitates the description of broadband and highly oscillatory laser fields containing multiple frequency components relevant to rovibrational transitions. The Fourier-feature vector is processed by a fully connected feedforward neural network containing multiple hidden layers and nonlinear activation functions. The final network output is constrained according to

\begin{equation}
E(t)
=
E_{\max}
\sin^2(\pi\tau)
\tanh\left[f_{\boldsymbol{\theta}}(\tau)\right],
\end{equation}
where $f_{\boldsymbol{\theta}}$ denotes the raw neural-network output and $E_{\max}$ is the maximum allowed field amplitude. The envelope function guarantees that the electric field vanishes smoothly at the beginning and end of the propagation interval, avoiding discontinuities and reducing nonphysical spectral broadening.

Although the proposed PINQC framework is inspired by the general philosophy of Physics-Informed Neural Networks (PINNs), its formulation differs from conventional PINN implementations. Rather than approximating the solution of the time-dependent Schrödinger equation with a neural network, the proposed approach parametrizes only the control field, while the quantum dynamics are computed explicitly using a differentiable split-operator propagator. This formulation preserves the underlying physical model and enables gradients to be propagated through the complete quantum evolution by automatic differentiation during the optimization process.

At each training epoch, the neural network generates a trial control field that is subsequently employed to propagate the rovibrational wave function according to the time-dependent Schrödinger equation. The propagation is performed using an explicit differentiable second-order split-operator method \cite{Feit1982}. For a sufficiently small time step $\Delta t$, the evolution operator is approximated by

\begin{equation}
U(t+\Delta t,t)
\approx
e^{-iH_0\Delta t/2\hbar}
e^{-iH_1(t)\Delta t/\hbar}
e^{-iH_0\Delta t/2\hbar}.
\end{equation}

This symmetric decomposition preserves the unitary structure of the quantum dynamics while improving numerical stability during long-time propagation. The optimization is performed through the minimization of the loss functional
\begin{equation}
\mathcal{L}(\boldsymbol{\theta})
=
\frac{1}{2}(1-F)
+
\alpha\int_0^{t_f}|E(t)|^2dt
+
\beta\int_0^{t_f}
\left|
\frac{dE(t)}{dt}
\right|^2dt.
\label{eq:loss_function}
\end{equation}
where the second term penalizes excessive pulse fluence and the third term imposes temporal smoothness on the optimized field. The parameters $\alpha$ and $\beta$ control the relative importance of these regularization terms. The gradient of the loss function with respect to the neural-network parameters,$\nabla_{\boldsymbol{\theta}}\mathcal{L}=\frac{\partial \mathcal{L}}{\partial \boldsymbol{\theta}}$, is computed automatically using reverse-mode automatic differentiation (backpropagation) implemented in PyTorch \cite{Paszke2019}. Physically, this gradient measures how infinitesimal modifications in the neural control policy affect the final rovibrational photoassociation probability.

The resulting methodology combines neural-network-based field generation, explicit differentiable quantum propagation, automatic differentiation, and gradient-based optimization into a unified optimization strategy for molecular quantum control. Unlike conventional quantum optimal-control approaches based on predefined pulse parametrization, the proposed framework learns the control field directly from the underlying quantum dynamics through end-to-end differentiable optimization. This strategy enables the exploration of complex laser-field structures while preserving the physical constraints imposed by the Schrödinger equation.

The neural-network parameters are optimized using the AdamW optimizer together with the automatic-differentiation capabilities provided by PyTorch. At each optimization epoch, the control field is updated through gradient-based optimization, and the complete quantum evolution is recomputed until convergence of the objective functional is achieved.

The performance of the proposed PINQC framework depends on both the neural-network architecture and the optimization hyperparameters. The hyperparameters employed throughout the present work were selected to ensure stable convergence and an accurate representation of the broadband laser fields required for rovibrational photoassociation. The main PINQC parameters are summarized in Table~\ref{tab:pinqc_parameters}. The hyperparameters reported in table were found empirically to provide stable convergence for all rovibrational calculations considered in the present work.

\begin{table}[htbp]
\caption{PINQC parameters used in the rovibrational optimization.}
\label{tab:pinqc_parameters}
\centering
\begin{tabular}{lc}
\hline\hline
Parameter & Value \\
\hline
$T$ & $120000$ a.u. \\
$N_t$ & $6000$ \\
$E_{\max}$ & $7.0$ \\
Hidden units & $384$ \\
Hidden layers & $3$ \\
Activation & Tanh \\
Number of Fourier features & $512$ \\
Optimizer & AdamW \\
Initial learning rate & $5\times10^{-6}$ \\
Epochs & $10000$ \\
Target fidelity & $0.99$ \\
$\alpha$ & $10^{-8}$ \\
$\beta$ & $10^{-10}$ \\
\hline\hline
\end{tabular}

\end{table}

\section{Results and Discussion}

In this section, we investigate the performance of the PINQC framework applied to the rovibrational photoassociation problem. We first analyze the convergence properties of the optimization procedure and the characteristics of the resulting control fields. Subsequently, we examine the rovibrational population dynamics induced by the optimized pulse and discuss the underlying photoassociation and stabilization mechanisms. Finally, we investigate the application of the PINQC framework to a rovibrational model containing larger rotational levels and compare its practical capabilities with those previously achieved in our conventional TBQCP calculations. The simulations were performed using the Morse parameters 
$D_e=0.1994~\mathrm{a.u.}$, $a=1.189~\mathrm{a.u.}$, and 
$r_e=1.821~\mathrm{a.u.}$. The initial continuum wave packet was modeled as a Gaussian wave packet centered at 
$r_0=25.3223~\mathrm{a.u.}$, with spatial width 
$d=15.3446~\mathrm{a.u.}$ and initial momentum
$p_0=-\sqrt{2\mu E_{\mathrm{initial}}-1/d^2}$, where 
$E_{\mathrm{initial}} = 9.18549 \times 10^{-4}\ \mathrm{a.u.}$. The transition dipole moment was modeled with $q=1.634$ and 
$r_d=1.13382~\mathrm{a.u.}$ \cite{Korolkov1996}. The rovibrational basis contained 
$N=172$ vibrational states and rotational levels up to $l_{\max}=4$ 
(or $l_{\max}=6$ in the extended calculations discussed separately in 
Section~III-D).

\subsection{Optimization performance}

We begin by analyzing the convergence properties of the proposed PINQC framework applied to the rovibrational photoassociation problem. The optimization objective is to maximize the population transferred from an initially continuum Gaussian wave packet into the vibrational ground-state level of the Morse potential. Because the control field is parametrized by a neural network and optimized through differentiable quantum propagation, successful convergence demonstrates that the proposed optimization strategy is capable of extracting physically meaningful gradient information directly from the underlying quantum dynamics.

\begin{figure}[htbp]
\centering
\includegraphics[width=0.75\linewidth]{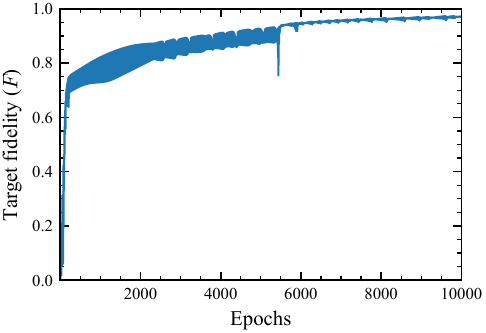}
\includegraphics[width=0.75\linewidth]{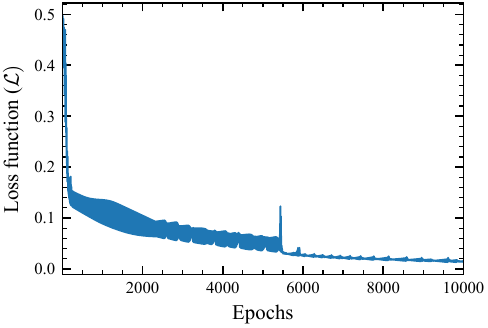}
\caption{
Evolution of the target fidelity (upper panel) and loss function (lower panel) during the PINQC optimization. The monotonic increase of the fidelity together with the decrease of the loss function demonstrates stable convergence of the differentiable optimization framework.
}
\label{fig:fidelity_loss}
\end{figure}

Figure~\ref{fig:fidelity_loss} presents the evolution of the target fidelity and the loss function throughout the optimization process. The fidelity increases smoothly during training while the loss function decreases monotonically, indicating stable convergence of the gradient-based optimization. No oscillatory behavior, convergence degradation, or numerical instabilities are observed during the optimization, demonstrating that the differentiable propagation and automatic-differentiation framework provide a robust optimization procedure even for the strongly coupled rovibrational dynamics considered in the present work.

The optimization converges to a final fidelity of approximately $0.98$, demonstrating that the PINQC framework successfully identifies laser fields capable of driving efficient continuum-to-bound population transfer. Achieving such a high fidelity is particularly significant because the optimization is performed over the complete rovibrational dynamics, where the laser field simultaneously couples multiple vibrational and rotational levels through the dipole interaction. The optimized control field therefore enables the formation of deeply bound molecules through efficient molecular photoassociation.

Although the fidelity evolution exhibits oscillations during training, its overall trend increases monotonically, as shown in Figure~\ref{fig:fidelity_loss}. This behavior indicates that the differentiable optimization remains numerically stable throughout the entire training procedure. Rather than relying on predefined pulse parametrization or iterative forward--backward update schemes, the neural-network representation progressively learns the control field directly from the underlying quantum dynamics, allowing increasingly efficient photoassociation strategies to emerge during the optimization process.

Having demonstrated the convergence and numerical stability of the optimization procedure, we next examine the optimized laser field and the corresponding rovibrational population dynamics that lead to efficient molecular photoassociation.

\subsection{Characteristics of the optimized laser field}

\begin{figure}[htbp]
\centering
\includegraphics[width=0.75\linewidth]{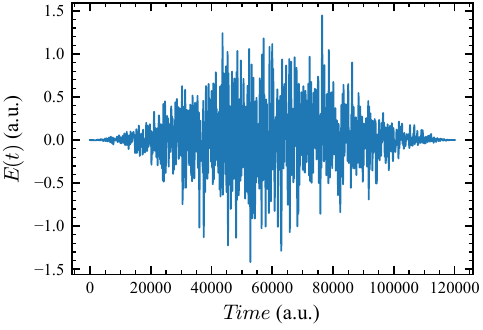}
\includegraphics[width=0.75\linewidth]{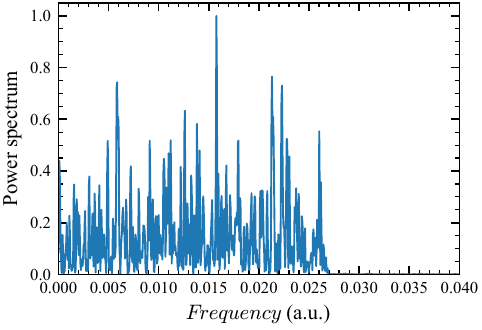}
\caption{
Optimized laser field generated by the PINQC framework (upper panel) and its corresponding power spectrum (lower panel). The broadband spectral distribution reflects the multichannel nature of the rovibrational photoassociation process, indicating that efficient continuum-to-bound population transfer is achieved through the cooperative action of multiple rovibrational transition pathways.
}
\label{fig:optimized_field}
\end{figure}

Having established the stable convergence of the optimization procedure, we now examine the characteristics of the optimized laser field generated by the PINQC framework. Figure~\ref{fig:optimized_field} presents both the temporal profile of the optimized pulse and its corresponding power spectrum. Rather than exhibiting a simple monochromatic or weakly modulated structure, the optimized field develops a highly structured temporal profile containing oscillatory components distributed over a broad range of frequencies. Such complexity reflects the multichannel nature of the rovibrational photoassociation problem, in which the laser pulse must simultaneously couple continuum-like states, weakly bound vibrational levels, and multiple rotational levels connected through dipole-allowed transitions.

Unlike conventional quantum optimal-control approaches, in which a user-defined initial laser pulse is iteratively optimized, the PINQC framework optimizes the parameters of a neural-network representation of the control field. The laser pulse is therefore generated by the neural network itself and progressively refined during training through automatic differentiation, allowing its temporal profile to emerge naturally from the underlying quantum dynamics rather than from a predefined analytical pulse parametrization.

The corresponding power spectrum further illustrates the complexity of the optimized control strategy. Instead of exhibiting a single dominant resonance, the spectrum displays a broad distribution of frequency components spanning the rovibrational transitions involved in the dynamics. This broadband behavior indicates that efficient photoassociation is achieved through the cooperative action of multiple transition pathways rather than through isolated resonant excitations. The optimized pulse therefore simultaneously addresses multiple rovibrational transitions required for the rovibrational photoassociation process considered here.

The emergence of this broadband spectral structure is particularly significant because it is not imposed \emph{a priori} by the pulse parametrization. Instead, it arises naturally during the optimization as the neural network progressively identifies the spectral components required to maximize the continuum-to-bound population transfer. This observation demonstrates that the differentiable optimization procedure automatically discovers physically meaningful laser-field structures directly from the quantum dynamics without relying on externally designed pulse shapes.

While the optimized laser field provides insight into the spectral resources required for efficient photoassociation, a more complete understanding of the control mechanism requires analyzing how the molecular population evolves throughout the optimization. We therefore next examine the time-dependent rovibrational population dynamics induced by the optimized pulse.

\subsection{Rovibrational photoassociation dynamics}

\begin{figure}[htbp]
\centering
\includegraphics[width=0.75\linewidth]{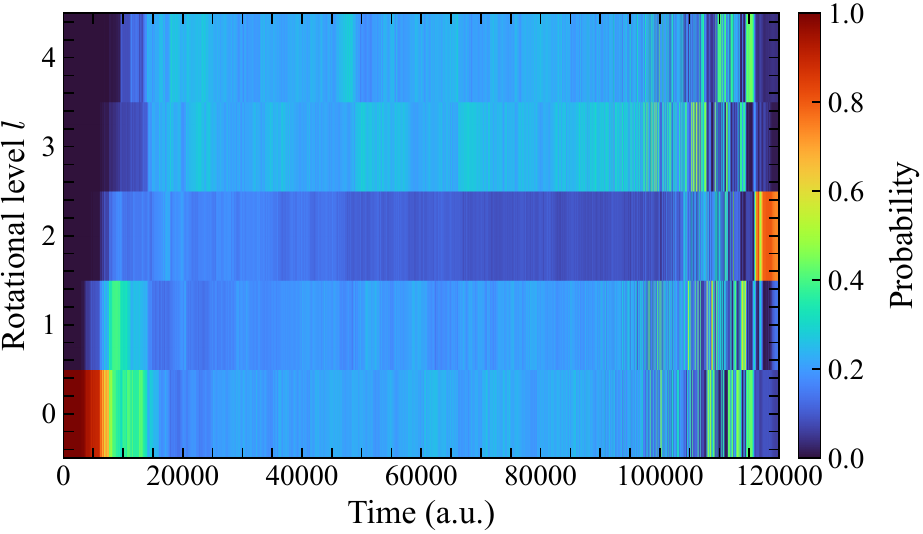}
\caption{
Time evolution of the rotational populations summed over all vibrational states. The optimized pulse redistributes population among multiple rotational levels through dipole-allowed transitions.
}
\label{fig:rotational_population}
\end{figure}

To understand the physical mechanisms underlying the optimization process, we now examine the rovibrational population dynamics induced by the optimized control field. Figure~\ref{fig:rotational_population} displays the time evolution of the rotational populations summed over all vibrational states,

\begin{equation}
P_l(t)=\sum_v P_{v,l}(t).
\end{equation}

Initially, the population is entirely localized in the rotational ground level ($l=0$), reflecting the preparation of the initial Gaussian wave packet in the spherical harmonic state $Y_0^0(\theta)$. As the optimized laser field interacts with the molecular system, dipole-induced transitions satisfying the selection rule $\Delta l=\pm1$ progressively redistribute population among neighboring rotational levels.

The resulting dynamics reveals substantial rotational mixing throughout the propagation, demonstrating that efficient photoassociation involves the cooperative participation of multiple coupled rotational levels rather than a single dominant transition pathway. At the end of the optimization, the rotational population is distributed over all accessible rotational states, with final populations of approximately $P_{l=0}=0.06$, $P_{l=1}=0.15$, $P_{l=2}=0.73$, $P_{l=3}=0.01$, and $P_{l=4}=0.04$. This distribution shows that significant population transfer occurs not only within the lowest rotational states but also across higher rotational levels, indicating that rotational redistribution plays an active role in the stabilization process. It is important to emphasize that this rotational population distribution is not prescribed by the optimization objective. Since the fidelity depends only on the total population accumulated in the vibrational ground-state manifold, the PINQC framework is free to exploit any rotational pathway that contributes to efficient continuum-to-bound transfer. The observed rotational redistribution therefore provides direct insight into the multichannel mechanisms selected by the differentiable optimization process.

\begin{figure*}[t]
 \centering
 \begin{subfigure}[h]{0.43\linewidth}
     \centering
     \includegraphics[width=\textwidth]{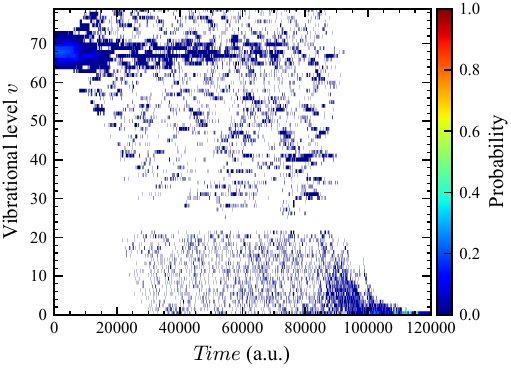}
     \caption{}
     \label{Fig4a}
 \end{subfigure}
 \hfill
 \begin{subfigure}[h]{0.43\linewidth}
     \centering
     \includegraphics[width=\textwidth]{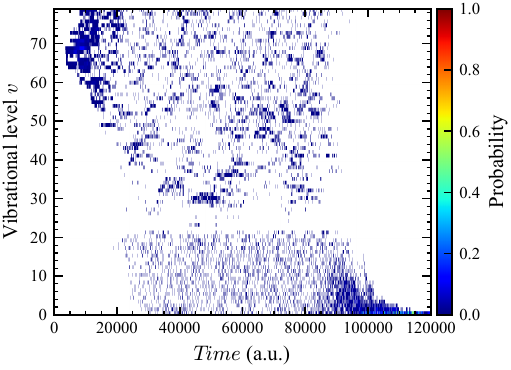}
     \caption{}
     \label{Fig4b}
 \end{subfigure}
 \hfill
 \begin{subfigure}[h]{0.43\linewidth}
     \centering
     \includegraphics[width=\textwidth]{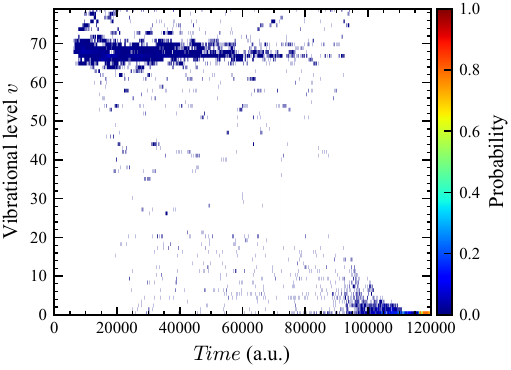}
     \caption{}
     \label{Fig4c}
 \end{subfigure}
 \hfill
 \begin{subfigure}[h]{0.43\linewidth}
     \centering
     \includegraphics[width=\textwidth]{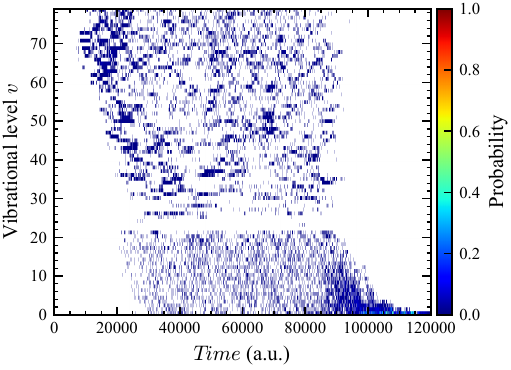}
     \caption{}
     \label{Fig4d}
 \end{subfigure}
  \hfill
 \begin{subfigure}[h]{0.43\linewidth}
     \centering
     \includegraphics[width=\textwidth]{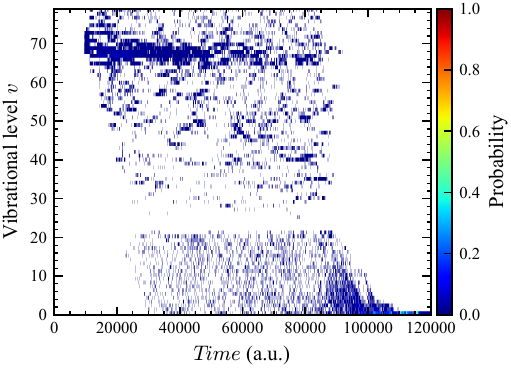}
     \caption{}
     \label{Fig4e}
 \end{subfigure}
    \caption{Time evolution of the vibrational populations for rotational levels: (a) $l=0$. (b) $l=1$. (c) $l=2$. (d) $l=3$. (e) $l=4$. The optimized field drives population from continuum-like states toward deeply bound vibrational levels through a multistep rovibrational stabilization process.}
    \label{Fig9}
\label{fig:rovib_maps}
\end{figure*}

To obtain a more detailed picture of the photoassociation mechanism, Figure~\ref{fig:rovib_maps} presents the time evolution of the vibrational populations associated with individual rotational levels. Each panel corresponds to a fixed rotational quantum number and illustrates how the population evolves among the vibrational states during the interaction with the optimized laser pulse.

The population maps reveal a clear flow from highly excited continuum-like states toward deeply bound vibrational levels. The optimized pulse initially promotes the capture of the incoming wave packet into weakly bound states located near the dissociation threshold. Subsequently, population is transferred through a sequence of intermediate rovibrational states, producing a gradual accumulation in lower vibrational levels and, ultimately, in the vibrational ground-state level.

This behavior is consistent with a sequential capture-and-stabilization mechanism. Rather than directly populating the target state, the optimized field exploits a network of intermediate rovibrational transitions that efficiently guide the population from the continuum into deeply bound molecular states. Significant transient population transfer among intermediate levels is observed throughout the propagation, indicating that the photoassociation process proceeds through multiple competing pathways distributed across the rovibrational Hilbert space.

The strongest stabilization occurs in the lower rotational levels, where a pronounced accumulation of population in the vibrational ground state ($v=0$) is observed at long propagation times. Nevertheless, higher rotational levels also participate actively in the transfer process, demonstrating that rotational redistribution plays an important role in the overall control strategy. These results show that the PINQC framework successfully identifies physically relevant multichannel pathways capable of driving efficient continuum-to-bound population transfer and population trapping in the vibrational ground-state level.

\subsection{Extension to a larger rovibrational model}

In our previous study of molecular photoassociation based on the Two-Point Boundary-Value Quantum Control Paradigm (TBQCP) \cite{ForlevesiTBQCP}, efficient continuum-to-bound population transfer was successfully achieved for the reference rovibrational model. However, attempts to extend the calculations to larger rotational levels resulted in rapidly increasing computational cost, mainly due to the repeated forward and backward propagations required by the optimization procedure. In practice, these computational limitations restricted the calculations to rotational levels of approximately $l_{\max}=4$, motivating the investigation of alternative optimization strategies capable of treating more complex rovibrational models.

To assess the capabilities of the proposed PINQC framework, additional calculations were performed including rotational levels up to $l_{\max}=6$, while maintaining the same vibrational basis adopted throughout the previous calculations. The inclusion of two additional rotational levels substantially enlarges the rovibrational Hilbert space and increases the number of dipole-coupled transition pathways participating in the photoassociation dynamics.

\begin{figure}[htbp]
\centering
\includegraphics[width=0.75\linewidth]{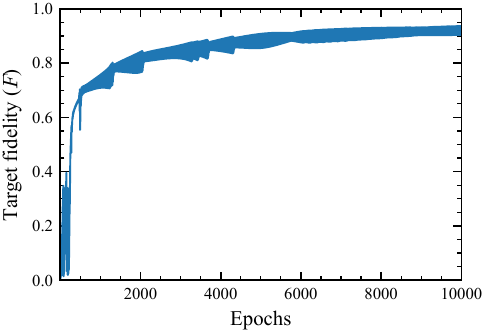}
\includegraphics[width=0.75\linewidth]{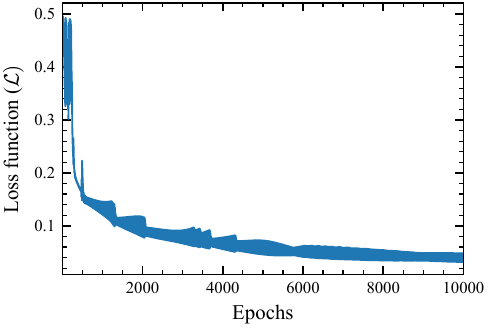}
\caption{
Evolution of the target fidelity (upper panel) and loss function (lower panel) during the PINQC optimization for the  rovibrational model including rotational levels up to $l_{\max}=6$. Stable convergence is maintained throughout the optimization despite the increased size of the rovibrational Hilbert space.
}
\label{fig:l6_fidelity}
\end{figure}

Figure~\ref{fig:l6_fidelity} presents the evolution of the target fidelity and the loss function during the optimization of the rovibrational model. The optimization converges to a final fidelity of approximately 0.95. Despite the increased dimensionality of the molecular system, the overall fidelity exhibits a monotonic upward trend throughout the training procedure, indicating the numerical stability of the differentiable optimization. The convergence behavior is qualitatively similar to that obtained for the reference model, demonstrating that the proposed PINQC framework successfully adapts to the larger rovibrational calculation.

\begin{figure}[htbp]
\centering
\includegraphics[width=0.75\linewidth]{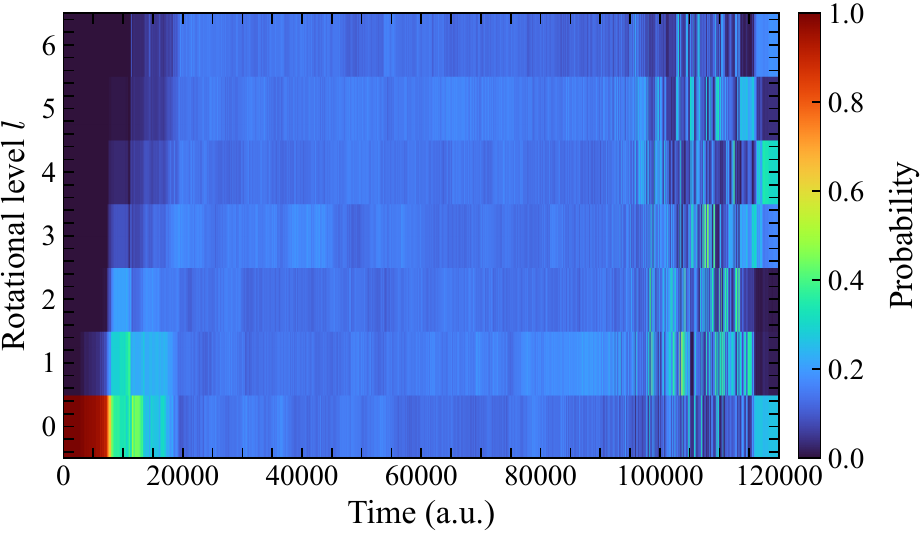}
\caption{
Time evolution of the rotational populations for the rovibrational model including rotational levels up to $l_{\max}=6$. The optimized pulse naturally explores the additional rotational pathways available in the enlarged rovibrational model.
}
\label{fig:l6_population}
\end{figure}

The corresponding rotational population dynamics are shown in Figure~\ref{fig:l6_population}. The optimized laser pulse simultaneously couples seven rotational levels ($l=0,\ldots,6$), producing a richer redistribution of rotational population than that observed for the $l_{\max}=4$ calculation. At the end of the propagation, the rotational populations are distributed as $P_{l=0}=0.27$, $P_{l=1}=0.02$, $P_{l=2}=0.01$, $P_{l=3}=0.17$, $P_{l=4}=0.31$, $P_{l=5}=0.04$, and $P_{l=6}=0.18$. These results demonstrate that the optimized photoassociation process proceeds through a genuinely multichannel rovibrational dynamics, with no single rotational level dominating the final population distribution. It is important to emphasize that the optimization objective remains unchanged, namely the maximization of the total population accumulated in the vibrational ground-state manifold. The additional rotational levels therefore do not constitute new optimization targets; rather, they provide additional dynamical pathways that are automatically exploited by the PINQC framework to enhance continuum-to-bound population transfer.

The successful optimization of the rovibrational model constitutes the principal result of the present work. Compared with our previous TBQCP implementation, the proposed PINQC framework allowed the optimization of a molecular model containing a larger rotational level while preserving stable convergence and efficient continuum-to-bound population transfer. Although the present work is restricted to rotational levels up to $l_{\max}=6$, these results demonstrate that differentiable optimization provides a practical strategy for treating molecular photoassociation problems beyond those previously accessible in our conventional calculations.

The present results should therefore be regarded as an initial demonstration of the applicability of the PINQC framework to larger rovibrational models. A systematic investigation of its computational performance for increasingly large Hilbert spaces, including detailed analyses of execution time, memory consumption, convergence behavior, and larger rotational levels, will be the subject of future work.

\begin{table}[htbp]
\caption{Largest rovibrational models successfully optimized in the present work and in our previous TBQCP implementation.}
\label{tab:comparison}
\centering
\begin{tabular}{lcc}
\hline\hline
Method & Largest $l_{\max}$ & Status \\
\hline
TBQCP & $\approx4$ & Computational limit \\
PINQC & $6$ & Stable optimization \\
\hline\hline
\end{tabular}
\end{table}

\section{Conclusion}

In this work, we introduced a Physics-Informed Neural Quantum Control (PINQC) framework for the rovibrational photoassociation of a molecular Morse system. By combining neural-network-based laser-field generation, differentiable quantum propagation, and automatic differentiation, the proposed approach enables laser pulses to be optimized directly from the underlying quantum dynamics without requiring external training data. The optimized fields successfully transfer an initially Gaussian wave packet into the vibrational ground-state level while naturally inducing the rotational redistribution required for efficient molecular photoassociation.

The numerical simulations demonstrate that the proposed framework yields laser fields capable of driving continuum-to-bound population transfer through multiple rovibrational pathways. Throughout the optimization, population is first captured from the continuum into weakly bound molecular states and subsequently redistributed among coupled rotational levels before being stabilized in the vibrational ground-state level. Because the optimization functional depends only on the total vibrational population, the final rotational distribution emerges naturally from the optimization process, providing insight into the multichannel mechanisms exploited by the optimized control field.

A central result of the present work is the successful application of the PINQC framework to a rovibrational model containing rotational levels up to $l_{\max}=6$. Compared with our previous photoassociation calculations based on the TBQCP method, this represents the successful optimization of a molecular system containing a larger rotational level while maintaining stable convergence and efficient continuum-to-bound population transfer. These results demonstrate that differentiable optimization constitutes a practical approach for treating rovibrational models of increased complexity within the context of molecular photoassociation.

The present work should therefore be regarded as an initial proof of concept demonstrating the applicability of differentiable quantum-control methods to rovibrational models. Although the present calculations were limited to rotational levels up to $l_{\max}=6$, the successful optimization of the enlarged molecular system motivates further investigations involving increasingly complex rovibrational dynamics and larger Hilbert spaces.

Future work will focus on extending the present framework to substantially larger rotational levels, more realistic molecular models, and state-selective quantum-control objectives. In addition, systematic studies of computational performance, numerical robustness, memory requirements, and convergence behavior will provide a deeper understanding of both the capabilities and the limitations of differentiable optimization for large-scale molecular quantum-control problems. We expect that these developments will further establish physics-informed differentiable optimization as a valuable computational tool for molecular quantum control and coherent manipulation of increasingly complex quantum systems.

\begin{acknowledgments}
The authors acknowledge the Center for Scientific Computing (NCC/GridUNESP) of the São Paulo State University (UNESP) and the financial support by resources supplied by FAPESP-São Paulo Research Foundation (No. 2024/09015-5, No. 2019/14038-6, No. 2021/09519-5, and No. 2024/22593-8), and by the National Council for Scientific and Technolog ical Development (CNPq), Brazil (No. 301318/2019-0, No. 304398/2023-3).
\end{acknowledgments}

\bibliographystyle{apsrev4-2}
\bibliography{ref}{}

\end{document}